# Photonics of shungite quantum dots

**B.S. Razbirin[1], N.N Rozhkova[2], E.F. Sheka[3]**

[1]*Ioffe Physical-Technical Institute, RAS, Saint Petersburg, Russia*
[2]*Institute of Geology Karelian Research Centre RAS, Petrozavodsk, Russia*
[3]*Peoples' Friendship University of Russia, Moscow, Russia*

**Abstract.** Shungite quantum dots are associated with nanosize fragments of reduced graphene oxide similarly to synthetic graphene quantum dots thus forming a common class of GQDs. Colloidal dispersions of powdered shungite in water, carbon tetrachloride, and toluene form the ground for the GQD photonic peculiarities manifestation. Morphological study shows a steady trend of GQDs to form fractals and a drastic change in the colloids fractal structure caused by solvent was reliably established. Spectral study reveals a dual character of emitting centers: individual GQDs are responsible for the spectra position while fractal structure of GQD colloids provides high broadening of the spectra due to structural inhomogeneity of the colloidal dispersions and a peculiar dependence on excitation wavelength. For the first time, photoluminescence spectra of individual GQDs were observed in frozen toluene dispersions which pave the way for a theoretical treatment of GQD photonics.

## 1. Introduction

Originally, the term 'graphene quantum dot' (GQD) appeared in theoretical researches and was attributed to fragments limited in size, or domains, of a single-layer two-dimensional graphene crystal. The subject of the investigations concerned the quantum size effects, manifested in the spin [1, 2], electronic [3] and optical [4-9] properties of the fragments. GQDs turned out quite efficient fluorescent nanocarbons. Due to the luminescence stability, nanosecond lifetime, biocompatibility, low toxicity, and high water solubility, the GQDs are considered as excellent probes for high contrast bioimaging and biosensing applications. The latter stimulated the growth of interest in GQD, so that the question arose of their preparation. In response to this demand, appeared a lot of synthetic methods to produce GQDs, both 'top-down' and 'bottom-up'. The former concern such techniques as electrochemical ablation of graphite rod electrodes, chemical exfoliation from the graphite nanoparticles, chemical oxidation of candle soots, intensive cavitation field in a pressurised ultrasonic batch reactor for obtaining nanosize graphite further subjected to oxydation and reduction. Laser ablation of graphite and microwave assisted small molecule carbonization peresent 'bottom-up' techniques [10]. These and other methods are widely used in the current studies concerning GQD photoluminescence (PL) (see Refs. [10-13] and references therein) and are described in a number of reviews [14, 15]. However the GQD technology is still costly and time consuming due to which working on efficient techniques getting GQDs in grams-mass quantities and searching cheap raw materials, chemists have turned to natural sources of carbon in the form of carbon fibers [16] and coal [17]. Sometimes GQDs of 'natural origin' are presented as carbon quantum dots (CQDs) [10].

In spite of a large variety of techniques as well as difference in the starting materials, numerous studies exhibit a common nature of both GQDs and CQDs. The performed analysis of structure and chemical composition show that in all cases GQDs are a few layer stacks of reduced graphene oxide (rGO) sheets of 1-10 *nm* in size. There is only one observation when single-layer rGO domains were synthesized in a liquid medium using trialkylphenyl polymers that form three-dimensional pores, inside which is a synthesis of carbon condensed polycyclic molecules [18]. The studied rGO stacks differ by the number of layers and linear dimension of rGO sheets as well by chemical composition of the latter. Thus, quasi planar basal plane of the sheets without chemical addends are framed quite differently with respect to chemical units that terminate the sheets' edge atoms: C=O, C-C(COOH), C-OH, and C-H units are considered as main terminators. As shown, a specific distribution of the latter over the sheets' circumference depends on which synthetic method of the dot producing was used. In its turn, it determines the solubility of GQDs in water and other solvents as well as a strong dependence of the GQDs PL properties on the latter. Therefore, the size- and chemical-composition- dependences are the main features of the GQDs PL spectra.

Due to the lack of a bandgap, no optical luminescence is observed in pristine graphene. A bandgap, however, can be engineered into GQDs due to quantum confinement [19, 20] and chemical modification of the graphene edge [21]. Since the bandgap depends on size [22], shape [23], and fraction of the $sp^2$ domains [24], PL emission greatly depends on the nature and size of the extended $sp^2$ sites [25]. This explains a large dispersion of PL properties that are affected by the synthetic method in use. The latter results in a large polydispersity of synthetic GQDs' solution and accounts for excitation-dependent PL shape and position. Standardization of both size and chemical framing of GQDs is a very complicated problem and not so many successful results are known. Controlled synthesis was realized in the case of GDQs derived from carbon fibers [16], encapsulated and stabilized in zeolitic imidazolate framework nanocrystals [26], and under a particularly scrupulous keeping of the protocol of chemical exfoliation of graphite [11].

As mentioned earlier, the greatest hopes in the production of a grams-mass GQDs' material are pinned on the use of natural source of carbon. Impressive results of the carbon fibers- [16] and coal [17] - based studies look quite promising. A particular feature of the studies concerned non-graphitic origin of both carbons while the produced GQDs do not differ from those of graphite-origin. Nevertheless, the natural carbons in the present case serve as raw material for complex chemical synthetic technology of the GQDs production only. At the same time, the Nature, infinitely generous to the carbon was not impassive at this time as well. As if anticipating the need for quantum dots, the Nature has prepared a particular natural carbon known as shungite. Deposited exclusively in the Karelia area of Russia, the shungite was attributed to one of the carbon allotropes quite long ago. Its bright individuality, as expressed, in particular, in the total absence of any similarity to other natural carbon allotropes, put shungite in a special position and formed the basis of more than half a century of careful study of its properties. The presence of $sp^2$ domains was one the first observations. However, it took many years to establish the domains' nature. Stimulating by the current graphene chemistry, both highly extensive empirically and deeply understood theoretically, the shungite study received a new impetus which resulted in a new vision of this carbon [27, 28]. At present, shungite is considered as a multilevel fractal structure which is based on nanoscale rGO fragments with an average linear dimension of ~1*nm*. The fragments are grouped in 3-5 layer stacks, thus forming the second level stacked structure; the stacks form the third level globular structures of 5-7 *nm* in size while the latter aggregate forming large nanoparticles of 20-100 *nm*. Experimental evidence of the structure has been recently proved by a detailed HRTEM study [29] and one of the obtained images is shown in Fig. 1. According to the image, shungite looks like 'buckwheat porridge' with the average grain size of ~1$nm$ and mean distance between grains ~0.35$nm$. These data are well consistent with both X-ray and neutron diffraction studies [30, 31]. A convincing proof of the grains' attribution to rGO fragments as well as their chemical composition describing the atomic C:O:H ratio of 11:1:3 were obtained in the course of inelastic neutron scattering study [32]. The latter has been recently confirmed by X-ray microanalysis [29]. Next level structure elements related to stacks, globules, and large aggregates were reliably fixed as well [29].

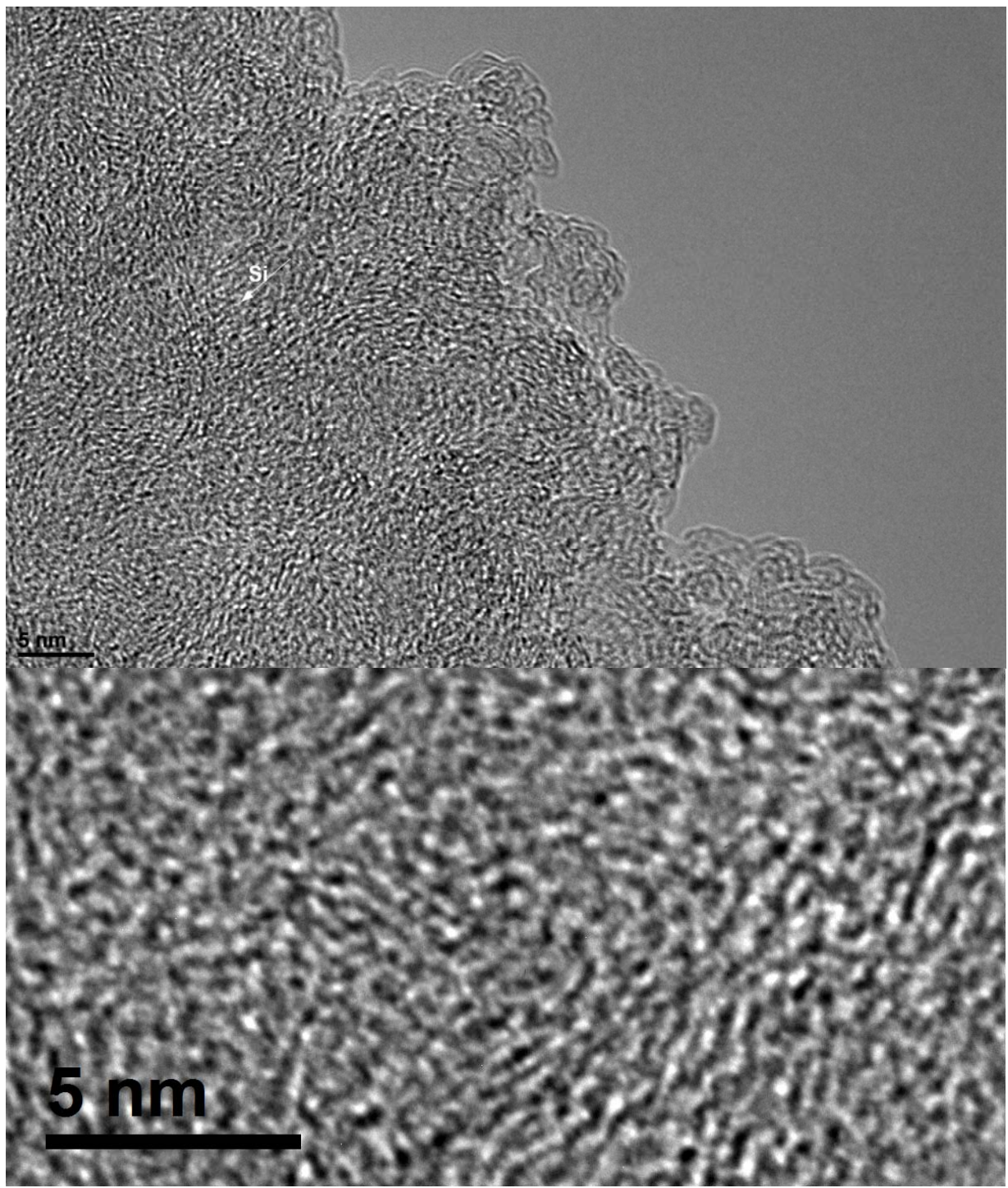

**Figure 1.** HRTEM image of a ~1μ shungite (Maksovo deposit) sample of thickness $d$ ($30 < d > 60nm$) at the FEI CM300UT/FEG instrument at two magnifications.

Coming back to GQDs, it should be stated that Fig.1 presents a picturesque image of the bodies related to shungite. Once stacked, globulised, and aggregated in solids, rGO fragments are readily dispersed in water and other solvents forming colloidal solutions with their own multilevel structures. From this viewpoint it is quite uncertain to speak about GQDs as a few layer stacks which is a nowadays vision of synthetic dots. Obviously, not stacks themselves but rGO fragments determine PL properties of GQDs. Apparently, would be better to attribute GQDs to these very fragments which will help to avoid problems caused by the fragment aggregation. On the other hand, such an approach opens a possibility to trace the influence of the fragment aggregation on PL properties by comparing spectra behavior of GQDs under different conditions. When applying to shungite GQDs, the approach turned out to be quite efficient which will be shown in the next Sections. Besides, a common nature of shungite and synthetic GQDs will greatly expand our knowledge on PL properties of GQDs.

**2. A prelude for photonics of shungite GQDs**

Optical spectroscopy and PL, in particular, has become the primary method of studying the spectral properties of the GQDs. The review [15] presents a synopsis of the general picture based on the study of synthetic GQDs, which can be presented by the following features.

- The position and intensity of the GQDs PL spectrum depend on the solvent; GQDs are readily soluble in water and many polar organic solvents; in the transition from tetrafuran to acetone, dimethyl formamide, dimethyl sulfide and water, maximum of the PL spectrum is gradually shifted from 475 to 515 nm, which evidently manifests the GQDs interaction with the solvent;
- The intensity of the PL depends on the pH of the solution: being very weak at low pH, it increases rapidly when the pH is from 2 to 12, wherein the shape of the spectrum does not change;
- An important feature of the GQDs PL is a variation in a wide range (from 2% to 46%) of its quantum yield; furthermore, this variation is associated not only with different ways of the GQD producing, but is typical of the samples prepared by the same procedure - the PL quantum yield varies with time after synthesis;
- Both GQDs absorption and PL spectrum shows the size dependence and shifts to longer wavelengths with increasing particle size;
- The PL spectrum depends on the excitation wavelength $\lambda_{exc}$, which is the result of inhomogeneous broadening of the absorption spectrum caused by a mixture of GQDs of different size and chemical composition; with increasing $\lambda_{exc}$, PL spectrum is shifted to longer wavelengths and its intensity is significantly reduced;
- The PL mechanism is still unclear, despite a large number of proposed models and active experimental studies.

Detailed description of these features with the presentation of their possible explanations and links to the relevant publications is given in Ref. [15].

As seen from the synopsis, optical spectroscopy of GQDs exhibits a complicated picture with many features. However, in spite of this diversity, common patterns can be identified that can be the basis of the GQDs spectral analysis, regardless of the method of their production. These general characteristics GQDs include the following: 1) structural inhomogeneity of GQDs solutions, better called dispersions; 2) low concentration limit that provides surveillance of the PL spectra; 3) dependence of the GQD PL spectrum on the solvent; 4) dependence of the GQD PL spectrum on the excitation light wavelength. It is these four circumstances that determine usual conditions under which the spectral analysis of complex polyatomic molecules is performed. Optimization of conditions, including primarily the choice of solvent and the experiments performance at low temperature, in many cases, led to good results, based on structural PL spectra (see, for example, the relevant research of fullerenes solutions [33-35]). In this chapter, we will show that implementation of this optimization for spectral analysis of the GQDs is quite successful.

## 3. Fractal Nature of the Object under Study

The GQD concept evidently implies a dispersed state of a number of nanosize rGO fragments.. Empirically, the state is provided by the fragments dissolution in a solvent. Once dissolved, the sheets unavoidably aggregate forming colloidal dispersions in water or other solvents. So far only aqueous dispersions of synthetic GQDs have been studied [14, 15]. In the case of shungite GQDs, two molecular solvent, namely, carbon tetrachloride and toluene were used as well when replacing water in the pristine dispersions. In each of these cases, the colloidal aggregates are the main object of the study. In spite of that so far there has not been any direct confirmation of their fractal structure, there are serious reasons to suppose that it is an obvious reality. Actually, first, the sheets formation occurred under conditions that unavoidably involve elements of randomness in the course of both laboratory chemical reactions and natural graphitization [28]; the latter concerns their size and shape and is clearly seen in Fig. 1. Second, the sheets structure certainly bears the stamp of polymers, for which fractal structure of aggregates in dilute dispersions has been convincingly proven (see Ref. [36] and references therein).

As shown in Ref. [36], the structure of colloidal aggregates is highly sensitive to the solvent around, the temperature of the aggregates formation, as well as other external actions such as mechanical stress, magnetic and electric field. In addition to previously discussed, this fact imposes extra restrictions on the definition of quantum dots of colloidal dispersions at the structural level and strengthen previous suggestion of attributing GQDs to rGO fragments. In our case of the GQDs of different origin, the situation is even more complicated since the aggregation of synthetic (Sy) and shungite (Sh) rGO fragments occurred under different external conditions. In view of this, it must be assumed that rGO-Sy and rGO-Sh aggregates of not only different, but the same solvent dispersions are quite different. Addressing spectral behavior of the dispersions, we should expect an obvious generality provided by the common nature of GQDs, but simultaneously complicated by the difference in packing of the dots in the different-solvent dispersions. The latter study concerned mainly the rGO-Sh dispersions [37, 38] that will be considered in detail below.

## 4. rGO-Sh Aqueous Dispersions

rGO-Sh aqueous dispersions were obtained by sonication of the pristine shungite powder [39]. The size-distribution characteristic profile of rGO-Sh aggregates is shown in Fig. 2a. As can be seen, the average size of the aggregates is 54 nm, whereas the distribution is quite broad and characterized by full-half-width-maximum (FHWM) of 26 nm. Thus, the resulting colloids are significantly inhomogeneous. The inhomogeneity obviously concerns both size and shape (and, consequently, chemical composition) of the basic rGO-Sh fragments and, consequently, GQDs. The structure of the carbon condensate formed after water evaporation from the dispersion droplets on a glass substrate is shown in the insert. As seen in the figure, the condensate is of fractal structure formed by large aggregates, the shape of which is close to spherical.

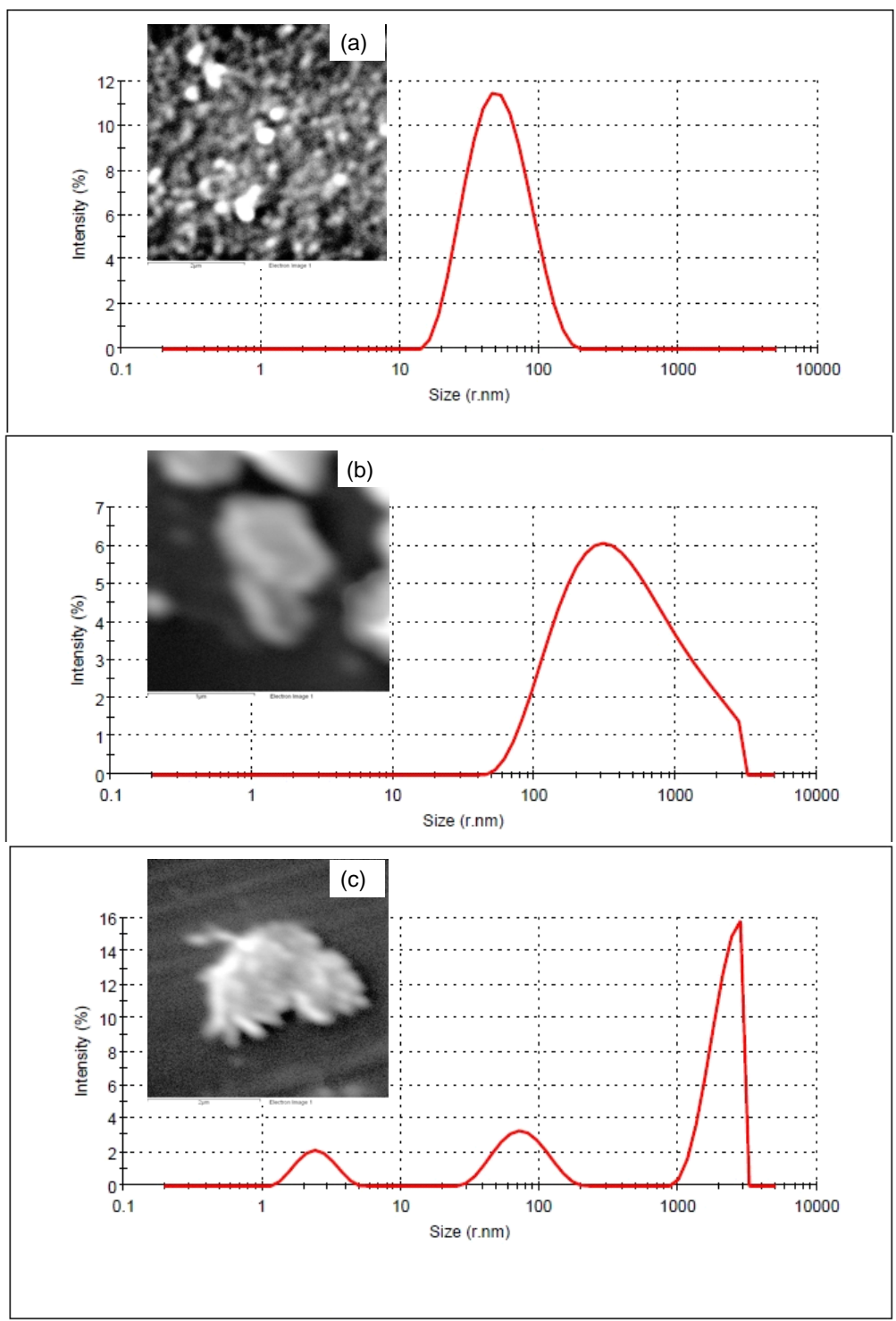

**Figure 2.** Size-distribution profile of colloidal aggregates of shungite in different dispersions. (a) Water; (b) Carbon tetrachloride; (c) Toluene. Carbon concentrations ~0.1 mg/ml. Inserts are SEM images of the dispersion condensate on glass substrate: Scale bar 2 μm.

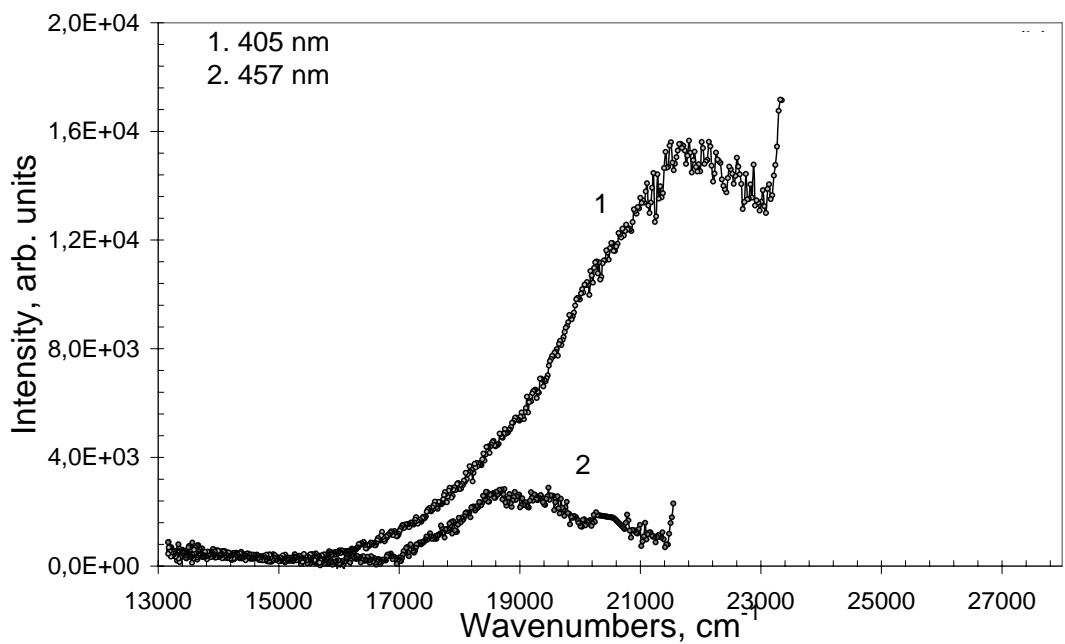

**Figure 3.** Photoluminescence spectra of shungite water dispersions at 290K after subtraction of Raman scattering of water. Numerals mark excitation wavelengths.

Emission spectra of rGO-Sh aqueous dispersions consist of PL and Raman spectrum (RS) of water at the fundamental frequency of the O-H stretching vibrations of ~3400 cm$^{-1}$[37]. Figure 3 shows the PL spectra excited at $\lambda_{exc}$ 405 and 457 nm after RS subtraction. The spectra present blue and green components of the GQD PL, both broad and bell-shaped, that are characteristic for the PL spectra of rGO-Sy aqueous dispersions (see Ref. [15]). In spite of large width of the PL spectra, their position in the same spectral region for both rGO-Sy and rGO-Sh aqueous dispersions evidences a common nature of emitting GQDs. The spectra are significantly overlapped with the absorption ones, which usually points to the presence of the inhomogeneous broadening. As a consequence, different $\lambda_{exc}$ provide a selective excitation of different sets of emitting centers. In the case of rGO-Sy dispersions, this feature was directly demonstrated by different fluorescence images (Fluoromax 4 (Horiba Scientific)) of the rGO-Sy colloids isolated in polymer film in the PL light excited by different $\lambda_{exc}$ showing the fluorescence originates from separate particles [12]. Figure 4 shows a similar picture for a drop of aqueous rGO-Sh colloids deposited on a glass at room temperature. As seen in the figure, at increasing $\lambda_{exc}$ blue emitting centers are complemented by green emitting ones while substituted with different red emitters. A variety of rGO-Sh fragments is clearly vivid in Fig. 1. Unfortunately, large width of PL spectra does not allow exhibiting those spectral details that might speak about aggregated structure of GQDs.

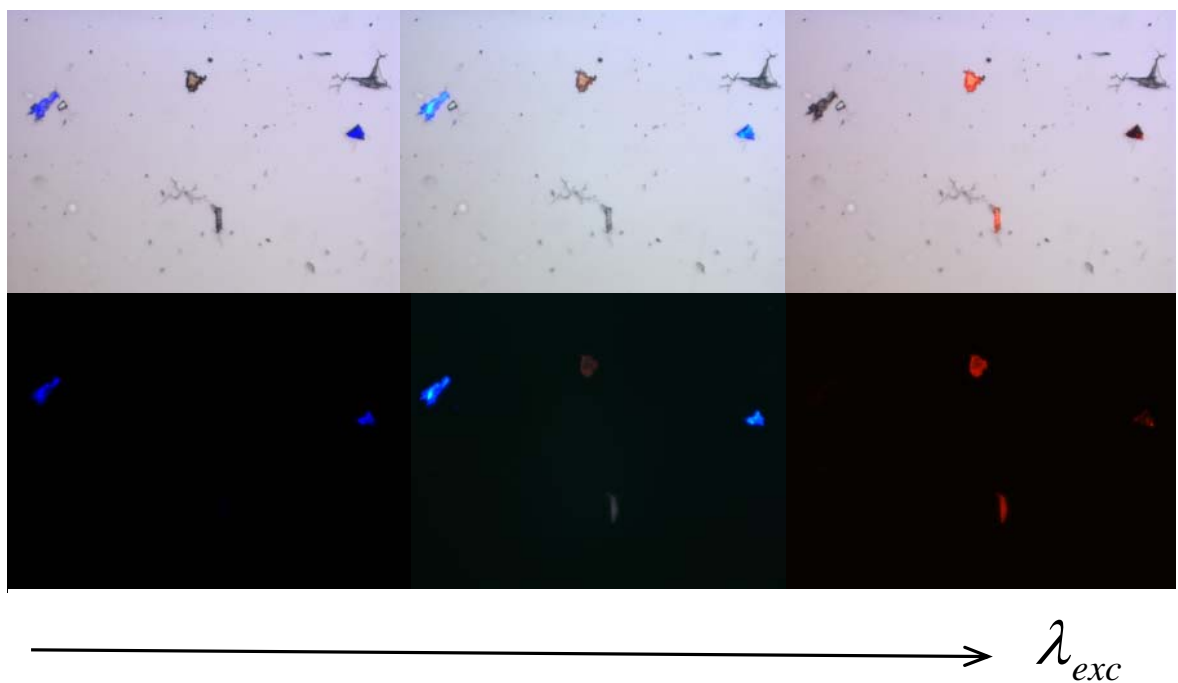

$\lambda_{exc}$

**Figure 4.** Fluorescence bright-field (top) and dark-field (bottom) images (Fluoromax 4 (Horiba Scientific)) of shungite aggregates deposited from water dispersion on a glass substrate excited by different wave lengths.

## 5. rGO-Sh Dispersions in Organic Solvents

Traditionally, the best way to overcome difficulties caused by inhomogeneous broadening of optical spectra of complex molecules is the use of their dispersions in frozen crystalline matrices. The choice of solvent is highly important. As known, water is a 'bad' solvent since the absorption and emission spectra of dissolved large organic molecules usually are broadband and unstructured. In contrast, frozen solutions of complex organic molecules, including, say, fullerenes [33-35], in carbon tetrachloride (CTC) or toluene, in some cases provide a reliable monitoring of fine-structured spectra of individual molecules (Shpolskii's effect [41]). Detection of PL structural spectra or structural components of broad PL spectra not only simplify spectral analysis but indicate the dispersing of emitting centers into individual molecules. It is this fact that was the basis of the solvent choice when studying spectral properties of shungite GQDs [37, 38].

Organic rGO-Sh dispersions were prepared from the pristine aqueous dispersions in the course of sequential replacement of water by isopropyl alcohol first and then by carbon tetrachloride or toluene [30]. The morphology and spectral properties of these dispersions turned out to be quite different.

### 5.1. *rGO-Sh Dispersions in Carbon Tetrachloride*

When analyzing CTC-dispersions morphology, a drastic change in the size-distribution profiles of the dispersions aggregates in comparison with that one of the aqueous dispersions was the first highly important feature. The second feature concerns the high incertitude in the structure of the latter.

Thus, Fig. 2b presents a size-distribution profile related to one of CTC-dispersions alongside with the TEM image of agglomerates of the film obtained when drying the CTC-dispersion droplets on glass. Typical for the dispersion is to increase the average size of colloidal aggregate when water is substituted with CTC. Simultaneously, increases the scatter of sizes that is comparable with the size itself. The nearly spherical shape of aggregates in Fig. 2a is replaced by lamellar faceting, mostly characteristic of microcrystals. It is necessary to note the absence of small aggregates, which indicates a complete absence of individual GQDs in the dispersions. Therefore, the change in both size-distribution profiles and shape of the aggregates of the condensate evidences a strong influence of solvent on the aggregates' structure.

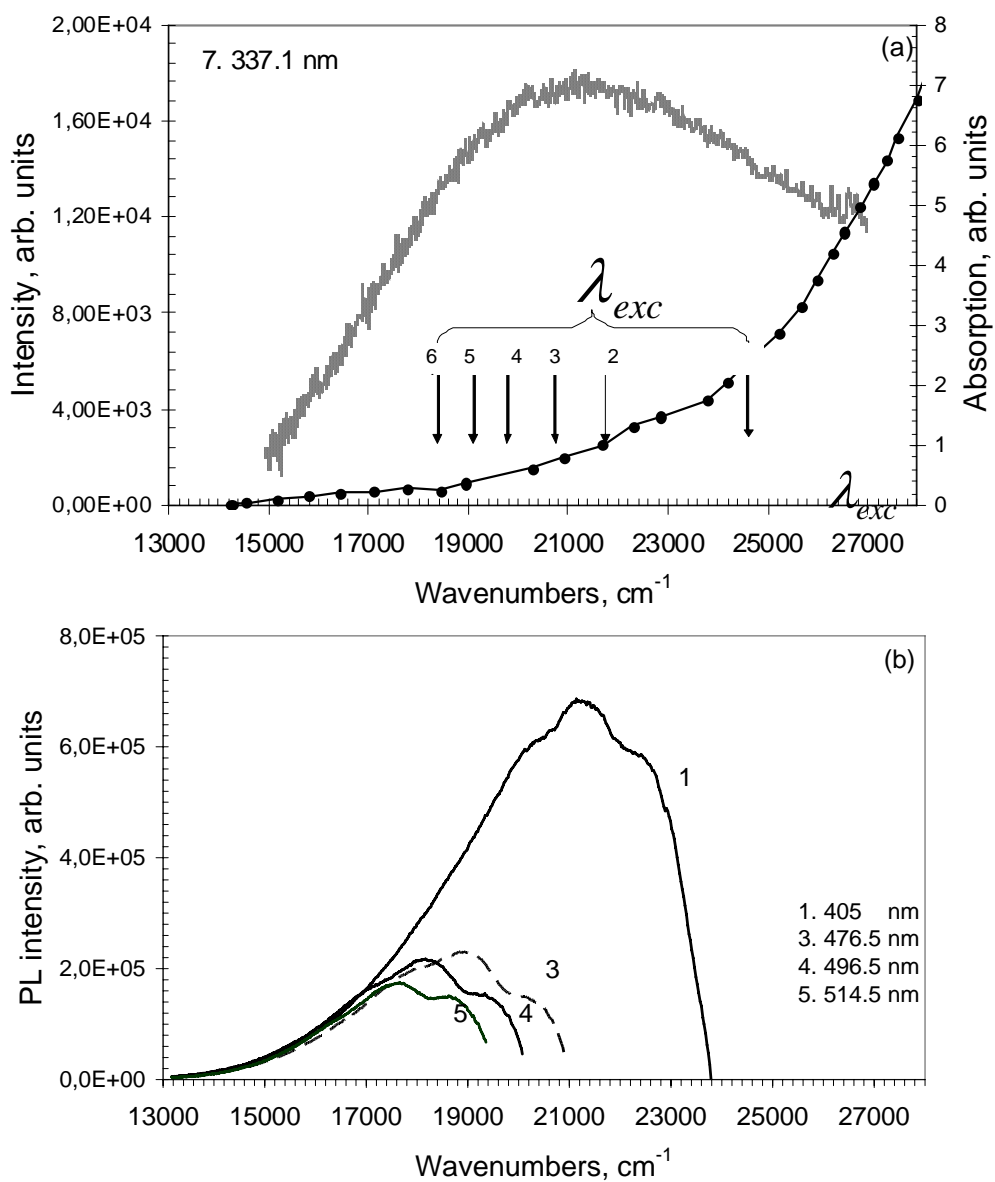

**Figure 5.** Photoluminescence (a and b) and absorption (a) spectra of shungite dispersion in carbon tetrachloride at 80K after background emission subtraction. Numerals mark excitation wavelengths.

The dispersion spectral features are well consistent with these findings. Figure 5 shows PL spectra of CTC-dispersion, morphological properties of which are similar to those shown in Fig. 2b. The dispersion is characterized by a wide absorption spectrum shown in Fig. 5a. Arrows in the figure indicate wave numbers $\lambda_{exc}^{-1}$ corresponding to laser lines. The UV excited PL spectrum in Fig. 5a is very broad and covers the region from 27000 to 15000 cm$^{-1}$ and overlaps with the absorption spectrum over the entire spectral range. Such a large overlapping evidences the formation of an ensemble of emitting centers, which differ in the probability of emission (absorption) at given wavelength. Indeed, successive PL excitation by laser lines 1, 3, 4 and 5 (see Fig. 5a) causes a significant modification of the PL spectra (Fig. 5b). The width of the spectra decreases as $\lambda_{exc}$ increases, the PL band maximum is shifted to longer wavelengths, and the spectrum intensity decreases. This is due to selective excitation of a certain group of centers which is typical for structurally disordered systems. To simplify further comparative analysis of the spectra obtained at different $\lambda_{exc}$, we shall denote them according to the excitation wavelength, namely: 405-, 476-, 496- spectrum, etc.

Comparing the discussed PL spectra at different excitations, note the following features:

- PL spectra obtained when excited in the region of overlapping of absorption and emission spectra in Fig. 5a, have more distinct structure than the 337-one but still evidencing a superpositioning character of the spectra;

- Intensity of the 405-spectrum is almost an order of magnitude higher than the intensity of the rest of the spectra.

As shown in Ref. [37], the features are typical for a wide range of dispersions obtained at different time. However, a comparative analysis of the PL spectra of different dispersions shows that the above-mentioned spectral regularities are sensitive to the CTC-dispersions structure and are directly related to the degree of structural inhomogeneity. Thus, the narrowing of the size-distribution profile undoubtedly causes narrowing of inhomogeneously broadened absorption and emission spectra. Unchanged in all the spectra is the predominance intensity of the 405-spectrum. The difference in the structural inhomogeneity of dispersions raises the question of their temporal stability. Spectral analysis showed that the spectra changed in the course of a long storage. Summarizing, the following conclusions can be made on the basis of spectral features of the rGO-Sh CTC-dispersions [38]:

1. None of fine-structured spectra similar to Shpolskii's spectra of organic molecules was observed in the low-temperature PL spectra of crystalline CTC-dispersions. This is consistent with the absence of small-size components in the size-distribution profiles of the relevant colloidal aggregates.
2. The PL spectra are broad and overlapping with the absorption spectrum over a wide spectral range. This fact testifies the inhomogeneous broadening of the spectra, which is the result of non-uniform distribution of the dispersions colloidal aggregates, confirmed by morphological measurements.
3. Selective excitation of emission spectra by different laser lines allows decomposing the total spectrum into components corresponding to the excitation of different groups of emitting centers. In this case, common to all the studied dispersions is the high intensity of the emission spectra excited at $\lambda_{exc}$ 405 and 457 nm.
4. The observed high sensitivity of PL spectra to the structural inhomogeneity of dispersions allows the use fluorescent spectral analysis as a method of tracking the process of the formation of primary dispersions and their aging over time.

5. The division of the GQD water dispersion spectra onto blue and green ones is not applicable to the spectra of the CTC dispersions. PL covers a large range from blue to red when $\lambda_{exc}$ increases.

*5.2. rGO-Sh Dispersion in Toluene*

The behavior of toluene rGO-Sh dispersions is more intricate from both morphological and spectral viewpoints. Basic GQDs of aqueous dispersions are awfully little soluble in toluene, thereby resulting toluene dispersions are essentially colorless due to low concentration of the solute. In addition, the low concentration makes the dispersion very sensitive to any change in both the content and structure of dispersions. This causes structural instability of dispersions which is manifested, in particular, in the time dependence of the relevant size-distribution profiles. Thus, the three-peak distribution of the initial toluene dispersion shown in Fig. 2c, is gradually replaced by a single-peaked at ~1*nm* for one to two hours. The last distribution does not change with time and represents the distribution of the solute in the supernatant.

By analogy with carbon tetrachloride, toluene causes a drastic change in the colloidal aggregates structure. However, if the carbon tetrachloride action can be attributed to the consolidation of the pristine colloids, the toluene results in quite opposite effect leading to their dispersing. Three-peak structure in Fig. 2c shows that, at the initial stage of water replacement by toluene, in resulting liquid medium there are three kinds of particles with average linear dimensions of about 2.5, 70 and 1100 nm. All the three sets are characterized by a wide dispersion. Large particles are seen in the electron microscope (see insert in Fig. 2c) as freaky sprawled fragments. Over time, these three entities are replaced by one with an average size of ~ 1*nm*. Thus, freshly produced dispersions containing GQD aggregates of varying complexity, turns into the dispersion of individual GQDs. This value is well consistent with the empirical value of ~1*nm* for the average size of GQDs in shungite accepted in Ref. [28] a seen in Fig. 1. The conversion of aqueous dispersion of aggregated GQDs into the colloidal dispersion of individual GQDs in toluene is a peculiar manifestation of the interaction of solvents with rGO. As for the graphene photonics, the obtained toluene dispersion has provided investigation of individual GQDs for the first time.

Figure 6 shows the PL spectra of colloidal dispersions of individual GQDs in toluene. The *brutto* experimental spectra, each of which is a superposition of the Raman spectrum of toluene and PL spectrum of the dispersion, are presented in Fig.6a. Note the clearly visible enhancement of Raman scattering of toluene in the 20000-17000 cm$^{-1}$ region. Figure 6b shows the PL spectra after subtracting Raman spectra. The spectra presented in the figure can be divided into three groups. The first group includes the 337-spectrum (7) that in the UV region is the PL spectrum, similar in shape to the UV PL spectrum of toluene, but shifted to longer wavelengths. This part of the spectrum should apparently be attributed to the PL of some impurities in toluene. The main contribution into the PL 337-spectrum in the region of 24000-17000 cm$^{-1}$ is associated with the emission of all GQDs available in the dispersion. This spectrum is broad and structureless, which apparently indicates the structural inhomogeneity of the GQD colloids.

PL 405- and 476-spectra (1 and 3) in the region of 23000-17000 cm$^{-1}$ should be attributed to the second group. Both spectra have clearly defined structure that is most clearly expressed in the 405-spectrum. The spectrum is characteristic of a complex molecule with allowed electronic transitions. Assuming that the maximum frequency at 22910 cm$^{-1}$ determines the position of pure electronic transition, the longer wavelength doublet at ~ 21560-21330 cm$^{-1}$ can be interpreted as vibronic satellites. The distance between the doublet peaks and the pure electronic band constitutes 1350-1580 cm$^{-1}$ that is consistent with the frequencies of totally symmetric vibrations of C-C graphene skeleton commonly observed in the Raman spectra. Similarly, two peaks of the much less intensive 476-spectrum, which are wider than in the previous case, are divided by the average frequency of 1490 cm$^{-1}$. PL 457-spectrum, shown in Fig. 6c (curve 2) is similar to spectra 1 and 3, in intensity closer to the 405-spectrum. All the three spectra are related to individuals rGO fragments albeit of different size that increases when going from 405-spectrum to 457- and 476-spectrum. All the spectra are positioned in blue-green region.

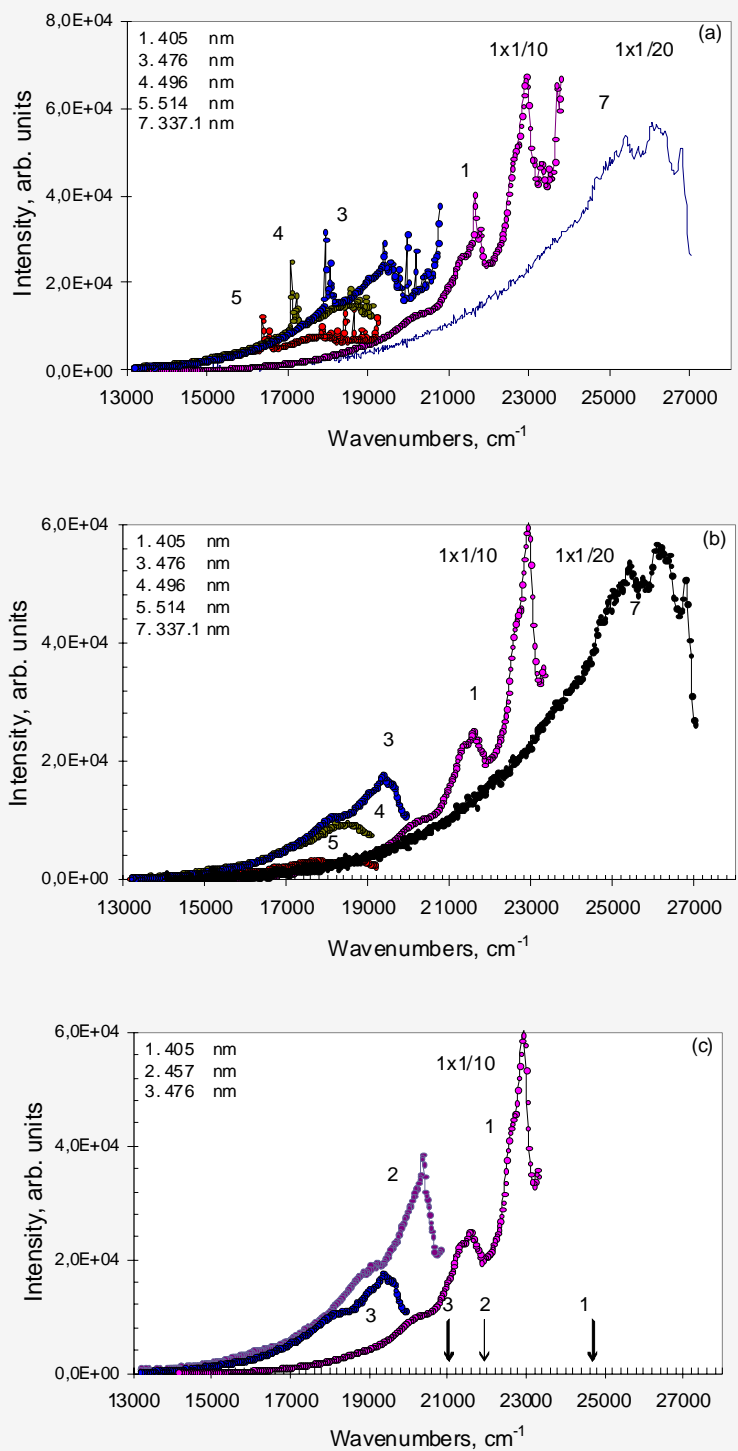

**Figure 5.** Photoluminescence spectra of shungite dispersions in carbon tetrachloride at 80K after background emission subtraction. A comparative view of 405- and 337-spectra (insert) of DC1 and DC2 dispersions (a); the same for spectra of DC2 dispersions at different excitations (b). Numerals mark excitation wavelengths.

The shape of 496- and 514-spectrum substantially differs from that of the second group spectra. Instead of the two peaks observed there a broad band is observed in both cases. This feature makes these spectra to attribute to the third group (red spectra) and to associate them with the appearance of not individual frozen GQDs but with their possible clusters (such as, say, dimeric homo- (GQD+GQD) and hetero- (GQD+toluene) structured charge transfer complexes and so forth) [38].

The conducted spectral studies of the rGO-Sh toluene dispersions confirmed once again the status of toluene as a good solvent and a good crystalline matrix, which allows for obtaining fine-structured spectra of individual complex molecules under conditions when in other solvents the molecules form fractals. This ability of toluene allowed for the first time to get the spectra of both individual GQDs and their small clusters. The finding represents the first reliable empirical basis for a comprehensive theoretical treatment of the spectra observed [42].

## 6. Discussion

As follows from the results presented above, rGO-Sh dispersions are colloidal dispersions regardless of the solvent, whether water, carbon tetrachloride or toluene. The dispersion colloids structure depends on the solvent and thereafter is substantially different. This issue deserves a special investigation. Thus, the replacement of water with carbon tetrachloride leads to multiple growth of the pristine colloids which promotes the formation of a quasi-crystalline image of the condensate structure. At present, the colloid detailed structure remains unclear. In contrast to carbon tetrachloride, toluene causes the decomposition of pristine colloids into individual rGO fragments. The last facts cast doubt on the possible direct link between the structure of the dispersions fractals and the elements of fractal structure of solid shungite or its post-treated condensate. The observed solvent-stimulated structural transformation is a consequence of the geometric peculiarities of fractals behavior in liquids [36]. The resulting spectral data can be the basis for further study of this effect.

The spectral behavior of the aqueous and CTC-dispersions with large colloids is quite similar, despite the significant difference in size and structure of the latter. Moreover, the features of the PL spectra of these dispersions practically replicate patterns that are typical for the aqueous rGO-Sy dispersions discussed in detail in Section 1. This allows one to conclude that one and the same structural element of the colloidal aggregates of both rGO-Sh dispersions and rGO-Sy one is responsible for the emission in spite of pronounced morphological difference of its packing in all these cases. According to the modern view on the shungite structure [28] and a common opinion on the origin of synthetic GQDs [14, 15], rGO sheets play the role thus representing GQDs of the rGO colloidal dispersions in all the cases.

Specific effects of toluene, which caused the decomposition of pristine particles into individual rGO fragments with succeeding embedding them into toluene crystalline matrix, allowed for the first time to obtain the PL spectrum of individual rGO fragments. Obviously, resulting fragments are of different size and shape, which determines the structural inhomogeneity of toluene dispersions. This feature of toluene dispersions is common with the other dispersions and explains the dependence of PL spectra on $\lambda_{exc}$ that is the main spectral feature of GQDs, both synthetic [14, 15] and of shungite origin.

The structural inhomogeneity of GQDs colloidal dispersions is caused by two main reasons, namely, internal and external. The internal reason concerns the uncertainty in the structure (size and shape) of the basic rGO fragments. Nanosize rGO basic structural elements of solid shungite are formed under the conditions of a serious competition of different processes [28], among which the most valuable are: 1) natural graphitization of carbon sediments, accompanied by a simultaneous oxidation of the graphene fragments and their reduction in water vapor; 2) the retention of water molecules in space between fragments and going out the water molecules from the space into the environment, and 3) the multilevel aggregation of rGO fragments providing the formation of a monolithic fractal structure of shungite. Naturally, that achieved balance between the kinetically-different-factor processes is significantly influenced by random effects, so that the rGO fragments of natural shungite, which survived during a Natural selection, are statistically averaged over a wide range of fragments that differ in size, shape, and chemical composition.

Obviously, the reverse procedure of the shungite dispersing in water is statistically also nonuniform with respect to colloidal aggregates so that there is a strong dependence of the dispersions on the technological protocol, which results in a change in the dispersion composition caused by slight protocol violations. Such, in a sense, a kinetic instability of dispersing, is the reason that the composition of colloidal aggregates can vary when water is displaced by other solvent. The discussed spectral features confirm these assumptions.

External reason is due to fractal structure of colloidal aggregates. The fractals themselves are highly inhomogeneous, moreover, they strongly depend on the solvent. The two reasons determine the feature of the GQD spectra in aqueous and CTC-dispersions while the first one dominates in the case of toluene dispersions. In view of this, photonics of GQDs has two faces, one of which is of rGO nature while the other concerns fractal packing of the rGO fragments. As follows from the presented in the chapter, spectra study is quite efficient in exhibiting this duality.

Thus, the structural PL spectra allow putting the question of identifying the interaction effect of dissolved rGO fragments with each other and with the solvent. Nanosize rGO fragments have high donor-and-acceptor properties (low ionization potential and high electron affinity) and can exhibit both donor and acceptor properties so that clusters of fragments (dimers, trimers, and so forth) are typical charge transfer complexes. Besides this, toluene is a good electron donor due to which it can form a charge-transfer complex with any rGO fragment, acting as an electron acceptor. The spectrum of electron-hole states of the complex, which depends on the distance between the molecules as well as on the initial parameters, is similar to the electron-hole spectrum of clusters of fullerenes $C_{60}$ themselves and with toluene [33-35], positioned by the energy in the region of 20000-17000 $cm^{-1}$. By analogy with nanophotonics of fullerene $C_{60}$ solutions, the enhancement of the RS of toluene is due to its superposition over the spectrum of electron-hole states, which follows from the theory of light amplification caused by nonlinear optical phenomena [43]. Additionally, the formation of rGO-toluene charge transfer complexes may promote the formation of stable chemical composites in the course of photochemical reactions [44] that might be responsible for the PL third-group spectra observed in toluene dispersions. Certainly, this assumption requires further theoretical and experimental investigation.

## 8. Conclusion

Photonics of shungite colloidal dispersions faces the problem that a large statistical inhomogeneity inherent in the quantum dot ensemble makes it difficult to interpret the results in details. Consequently, most important become common patterns that are observed on the background of this inhomogeneity. In the case of considered dispersions, the common patterns include, primarily, the dispersion PL in the visible region, which is characteristic for large molecules consisting of fused benzenoid rings. This made it possible to confirm the earlier findings that graphene-like structures of limited size, namely, rGO fragments are the basic structural elements for all the dispersions. The second feature concerns the dependence of the position and intensity of selective PL spectra on the exciting light wavelength $\lambda_{exc}$. This feature lies in the fact that regardless of the composition and solvent of dispersions the PL excitation at $\lambda_{exc}$ 405 and 457 nm provides the highest PL intensity while excitation at either longer or shorter wavelengths produces a much lesser intensity of the emission. The answer to this question must be sought in the calculated absorption and photoluminescence spectra of graphene quantum dots, which are attributed to nanoscale fragments of reduced graphene oxide [42].

**Acknowledgement**

The financial support provided by the Ministry of Science and High Education of the Russian Federation grant 2.8223.2013 and the Basic Research Program, RAS, Earth Sciences Section-5, is highly acknowledged. The authors are thankful to D.K. Nelson, A.N. Starukhin for a valuable assistance in performing spectral experiments as well as to E.A.Golubev and E.A.Suvorova for a kind permission to use one of their HRTEM images and to A.S.Goryunov for supplying with size-distribution profile measurements.